\documentclass[11pt]{article}

\PassOptionsToPackage{hyphens}{url}

\usepackage[preprint]{acl}

\usepackage{times}
\usepackage{latexsym}

\usepackage[T1]{fontenc}

\usepackage[utf8]{inputenc}
\usepackage{amsmath}
\usepackage{booktabs}
\usepackage{multirow}
\usepackage{float}
\usepackage{microtype}

\usepackage{inconsolata}

\usepackage{graphicx}
\usepackage{algorithm}
\usepackage{algpseudocode}

\author{Rong Wang \\
  Florida State University \\
  \texttt{rw25r@fsu.edu} \\\And
  Mulong Luo \\
  Florida State University \\
  \texttt{ml25bj@fsu.edu}}

\title{\textsc{ECHO}: Revealing Review-to-Submission Linkage in Anonymous Peer Review}

\begin{document}
\maketitle
\begin{abstract} 
Anonymous peer review protects reviewers from retaliation, yet reviewer anonymity may also be weakened by artifacts released through the review process itself. This risk is growing as many venues require submitting authors to review, creating substantial overlap between authors and reviewers within the same review cycle. We identify \emph{review-to-submission linkage}, where an anonymous review can be associated with the reviewer's own submission, substantially shrinking the reviewer anonymity set. To assess whether this linkage is practically exploitable, we develop \textsc{ECHO}, which combines technical-expertise and writing-style signals across reviews, papers, and rebuttals, and evaluate it on R2S-Bench, a ground-truth benchmark we construct. \textsc{ECHO} correctly identifies the reviewer's own submission in Top-1 for 34.6\% of reviews, in the Top-5 for 54.0\%, and in the Top-10 for 63.4\%, compared with 0.5\%, 2.6\%, and 5.3\% under random ranking, from the candidate submissions. With a stronger Stage-2 reranker, \textsc{ECHO} further reaches 56.4\% Top-1 and 66.4\% Top-10. These results demonstrate that review-to-submission linkage is feasible using only routinely released peer-review artifacts. Code and benchmark are available at \url{https://github.com/kaimu11111/ECHO-Revealing-Review-to-Submission-Linkage-in-Anonymous-Peer-Review}.
\end{abstract}

\begin{figure*}[t]
    \centering
    \vspace{-2mm}
    \includegraphics[width=0.85\textwidth]{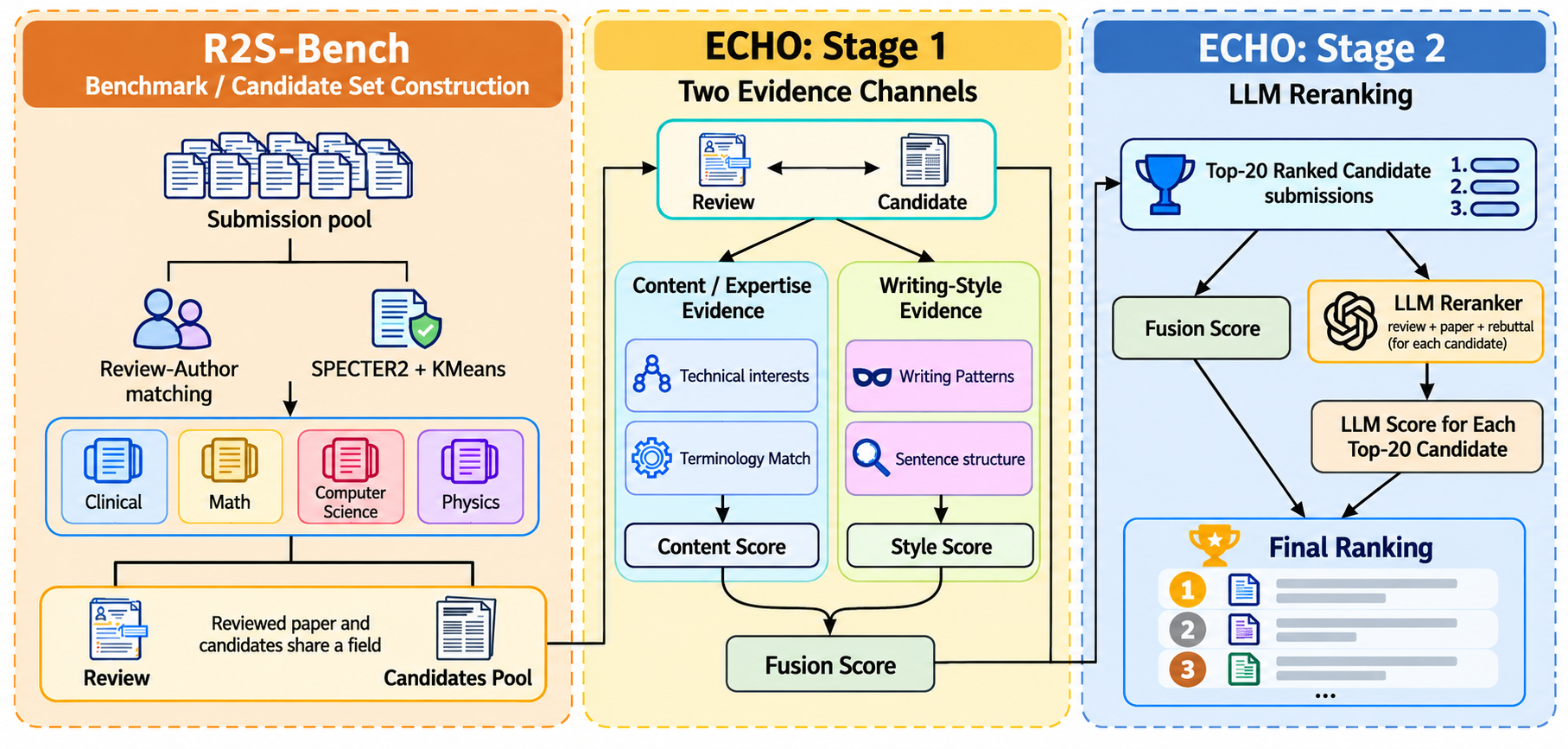}
    \vspace{-2mm}
    \caption{
    Overview of the review-to-submission linkage pipeline. The left panel shows benchmark and candidate-set construction, where reviewer--author matching establishes ground truth and SPECTER2-based research-area grouping defines a topically plausible candidate set. Given this candidate set, \textsc{ECHO} combines content/expertise and writing-style evidence to produce a fusion score, then reranks the top-20 candidates with an LLM and combines both scores to obtain the final ranking.
    }
    \label{fig:overview}
    \vspace{-2mm}
\end{figure*}
\section{Introduction}

Anonymous peer review protects the independence of scientific evaluation
by allowing reviewers to provide candid assessments without fear of
retaliation or strategic pressure. At the same time, many venues publicly
release anonymous submissions, reviews, and author responses during the
review process, making reviewer anonymity increasingly important.

The consequences of losing this protection became particularly visible during the ICLR 2026 review cycle, when a vulnerability in the OpenReview API exposed the otherwise anonymous identities of authors, reviewers, and area chairs associated with more than 10,000 submissions---approximately {45\%} of the conference \cite{iclr2026incident}. The incident was followed by reports of attempted collusion, harassment, intimidation, and bribery directed at reviewers, illustrating that reviewer anonymity is not merely a procedural convention but an important protection for the integrity of peer review.

The ICLR incident resulted from an explicit failure of the review platform. In this work, we ask a different question: \emph{can reviewer anonymity be substantially reduced even when the review system itself reveals no private information?} Removing names, affiliations, and other explicit identifiers does not necessarily remove all identity-related signals from scientific text. Prior work has shown that anonymized manuscripts retain information about their authors through citation patterns, writing style, topical interests, and other textual cues \cite{hill2003myth,payer2015what,bauersfeld2023cracking}. Similar signals also appear in reviews, where textual patterns can be linked to previously observed reviews or broader publication records \cite{nanavati2011herbert,seroussi2014authorship,zhang2026deanonymization}. These findings suggest that anonymity leakage may arise not only from private metadata or platform failures, but also from the textual artifacts routinely released during peer review.

We study a different linkage that arises from a common feature of modern peer review, where reviewers are often authors of other papers from the same venue. At large conferences, researchers who submit papers are routinely mandated to review other submissions in the same cycle. Consequently, an anonymous review may coexist with other publicly available artifacts written by the same researcher---the reviewer's own submission and, when available, its author response. Although reviews, papers, and rebuttals serve different purposes, they can reflect overlapping technical expertise, preferred terminology, recurring research interests, and characteristic writing patterns. This creates a cross-artifact linkage:
\[
\text{\textit{anonymous review}}
\;\longrightarrow\;
\text{\textit{reviewer's own manuscript}}.
\]

This linkage differs from prior work on manuscript and reviewer attribution. Existing approaches link anonymous manuscripts to authors or match anonymous reviews against previously observed reviews, publication histories, or other texts from the same writer \cite{hill2003myth,payer2015what,bauersfeld2023cracking,nanavati2011herbert,seroussi2014authorship,zhang2026deanonymization}. In contrast, our setting uses only artifacts routinely released during peer review and requires neither reviewer identity profiles nor previously labeled reviews from the target reviewer. Given an anonymous review and a set of plausible submissions, we ask whether this cross-artifact linkage is strong enough to be exploited in practice. Specifically, we test whether the reviewer's own submission can be consistently ranked among a small subset of plausible candidates using only information released through the normal peer-review process. Because each candidate is a submission rather than an identity profile, a successful linkage directly reduces the relevant anonymity set to that paper's authors. We refer to this threat as \emph{review-to-submission linkage}.

A key challenge is that a review reflects both the paper being evaluated and the reviewer who writes it. Much of the review's topical content is naturally determined by the reviewed submission, which can obscure signals associated with the reviewer's own research interests. As a result, simple semantic matching may favor papers similar to the reviewed work rather than the reviewer's own submission. The problem is further complicated by the fact that reviews, papers, and rebuttals are written for different purposes and exhibit different linguistic characteristics. We therefore combine complementary signals that capture both the reviewer's technical expertise and more persistent writing patterns across these heterogeneous artifacts.

Large-scale evaluation is difficult because double-blind venues do not reveal the ground-truth link between an anonymous review and the reviewer's own submission. We therefore construct R2S-Bench (Review-to-Submission Linkage Benchmark) from 11 journals on the F1000 open-review platform \cite{f1000research}, where reviews are publicly signed and articles are accompanied by author responses. Reviewer identities are used only to establish ground truth and are removed from all linkage inputs. The corpus contains 12,466 articles, 37,673 reviews, and 13,117 author responses, with 2,727 researchers matched as both reviewers and authors using ORCID or conservative name-and-affiliation matching. To approximate submissions available within the same review cycle, we group articles into eight research areas and construct topically and temporally plausible candidate sets, with each candidate representing a single submission rather than an author profile.

Our experiments show that publicly released peer-review artifacts can substantially reduce reviewer anonymity. Each review is ranked against a median of 414 candidate submissions from the same research area. To test whether this linkage can be exploited in practice, we develop \textsc{ECHO} (\textbf{E}xpertise and \textbf{C}ross-artifact \textbf{H}abits for \textbf{O}wnership), a review-to-submission linkage method that combines technical-expertise and writing-style evidence across reviews, papers, and rebuttals. Figure~\ref{fig:overview} provides an overview of the benchmark
construction and the \textsc{ECHO} linkage pipeline. \textsc{ECHO} ranks the reviewer's own submission in the top 5 for \textbf{54.0\%} of reviews and in the top 10 for \textbf{63.4\%}. These results show that meaningful anonymity reduction can arise from ordinary peer-review artifacts alone, without access to private metadata or the review platform. More broadly, they suggest that peer-review anonymity should be considered not only at the level of individual documents, but also in terms of the relationships among submissions, reviews, and author responses released across the review process.

We make three main contributions.
\begin{figure*}[t]
  \centering
  \includegraphics[width=\textwidth]{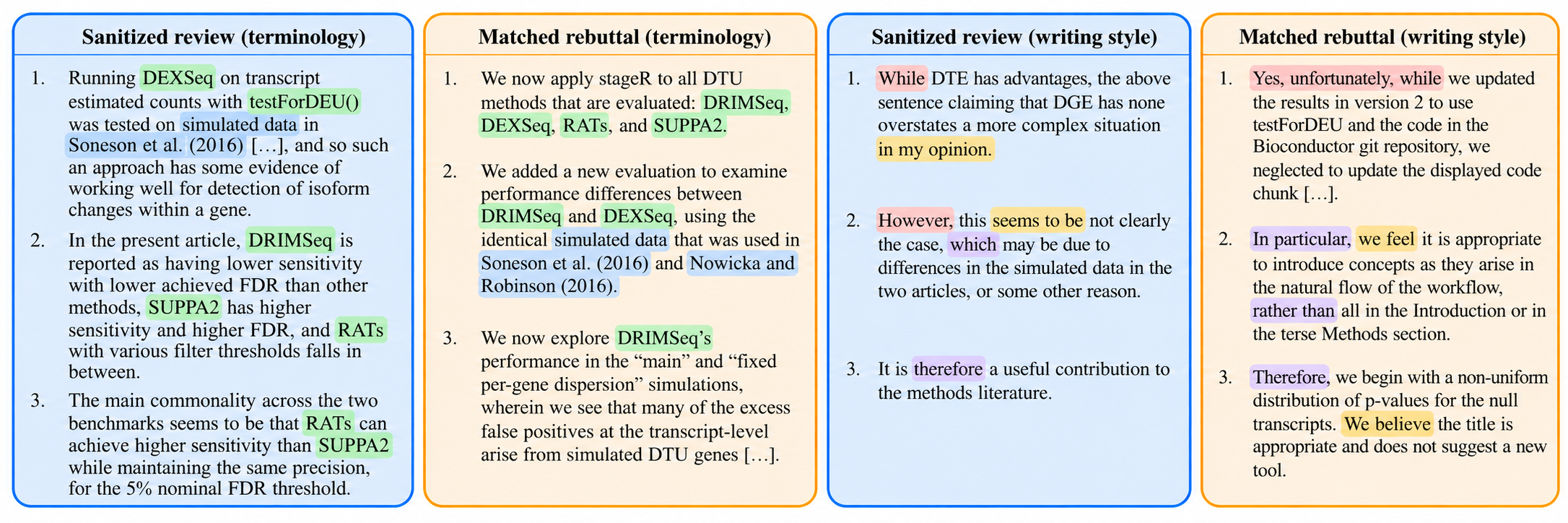}
  \caption{Examples of matched terminology and writing-style cues between a sanitized review and the matched rebuttal. Corresponding terms and phrases are highlighted in the same color.}
  \label{fig:example}
\end{figure*}
\begin{itemize}
    \item We identify \emph{review-to-submission linkage} as a privacy risk in double-blind peer review, where an anonymous review can be associated with the reviewer's own submission using only artifacts routinely released during the review process.

    \item We construct a large-scale ground-truth benchmark that makes this otherwise hidden linkage measurable, without relying on private platform metadata or previously labeled reviews from the target reviewer.

    \item We develop \textsc{ECHO}, which combines content and writing-style evidence to operationalize review-to-submission linkage, and empirically demonstrate that the linkage is practically feasible across hundreds of plausible candidate submissions, and the leakage persists even after AI-based rewriting suppresses stylistic cues.
\end{itemize}

\section{R2S-Bench: Review-to-Submission Linkage Benchmark}
\label{sec:benchmark}
Evaluating review-to-submission linkage requires ground-truth connections between reviews and submissions written by the same reviewer. Such relationships are unavailable at double-blind venues because reviewer identities are hidden.  We therefore construct a benchmark called R2S-Bench (Review-to-Submission Linkage Benchmark) from the F1000 open-research platform \cite{f1000research}, where reviews are publicly signed and articles are accompanied by author responses. We refer to these author responses as \emph{rebuttals} throughout the paper. Reviewer identities are used only to establish ground truth and are removed from all benchmark inputs.

\subsection{Corpus and Reviewer--Author Matching}

Our corpus spans 11 journals and contains 12,466 articles, 37,673 reviews,
and 13,117 rebuttals. We identify 2,727 reviewer--authors using ORCID or
conservative name-and-affiliation matching, excluding name-only matches.
Reviewer identities are replaced with pseudonymous identifiers before
benchmark construction. Full matching rules are provided in Appendix~\ref{app:identity}.

\subsection{Review Sanitization}

Because the original reviews are signed, we remove explicit identity cues
before using them as benchmark queries, including names, affiliations,
email addresses, ORCID identifiers, profile links, signatures, and platform
templates. A stricter version additionally removes publication identifiers
and obvious self-references. Manual inspection of 500 sanitized reviews
found no remaining direct identity cues. Full rules are provided in
Appendix~\ref{app:sanitize}. 

\subsection{Research Areas and Candidate Sets}

Reviewer assignment is typically driven by topical expertise and
paper--reviewer affinity
\citep{charlin2013tpms,thorn-jakobsen-rogers-2022-factors}.
We therefore construct candidate sets from topically related submissions
rather than selecting candidates randomly from the entire corpus.
Scientific document embeddings have been widely used to capture
paper-level semantic relatedness
\citep{cohan-etal-2020-specter,ostendorff-etal-2022-neighborhood}.
We encode the title and abstract of each paper using
SPECTER2~\citep{singh-etal-2023-scirepeval} and group the papers into
eight broad research areas with KMeans clustering.

Because our evaluation considers the full set of publicly released
review artifacts, we focus on venues where rebuttals are also available.
Based on this setting, for each anonymous review, candidate submissions
are drawn from the same research area and restricted to papers that fall
within the corresponding review-cycle time window and have an accompanying
rebuttal. We refer to this set as the \emph{submission pool}. We
conservatively exclude any other article whose authorship could plausibly
match the reviewer. Otherwise, more than one candidate could be a valid
review-to-submission link, making the ground-truth target ambiguous.

This produces topically plausible alternatives while avoiding ambiguous
negatives. Each candidate corresponds to a single submission. We never
combine multiple papers from the same researcher into an author profile,
since grouping multiple submissions by the same hidden author would itself
require identity information that is unavailable in double-blind review.
\subsection{Benchmark Instances}
Each instance in R2S-Bench consists of one sanitized anonymous review, a set of candidate submissions, and one ground-truth submission authored by the reviewer. Each candidate is represented by two artifacts from the same submission. The paper representation contains its title and abstract, while the rebuttal provides an additional sample of the authors' technical writing.

Rebuttals are cleaned before use by removing quoted review text, signatures, and explicit identity information. For the ground-truth submission, we retain the rebuttal written by the matched reviewer.

After applying the area, time, rebuttal, and ground-truth constraints, the
benchmark contains 298 instances from 256 reviewer--authors, with a median
of 414 candidates per instance. No explicit reviewer identity fields are
included. Table~\ref{tab:benchmark_summary} summarizes the benchmark, with additional
construction details in Appendix~\ref{app:identity}.

\begin{table}[t]
\centering
\small
\begin{tabular}{lr}
\toprule
\textbf{Statistic} & \textbf{Value} \\
\midrule
Articles & 12,466 \\
Reviews & 37,673 \\
Rebuttals & 13,117 \\
Reviewer--authors & 2,727 \\
Research areas & 8 \\
Benchmark instances & 298 \\
Unique reviewer--authors & 256 \\
Median candidates per instance & $\sim$414 \\
Candidate representation & paper + rebuttal \\
\bottomrule
\end{tabular}
\caption{Summary of the review-to-submission linkage benchmark.}
\label{tab:benchmark_summary}
\end{table}

\section{ECHO: Expertise and Cross-artifact Habits for Ownership}
\label{sec:method}

We describe ECHO, our method for review to submission linkage.
Given the candidate set $\mathcal{C}(q)$ of a benchmark instance, ECHO proceeds in two stages. As shown in Figure~\ref{fig:overview}, in Stage 1, a conditional-logit model fuses two complementary families of evidence: content similarity between the review and both candidate artifacts, and writing-style similarity between the review and the candidate rebuttal, to produce an initial ranking. In Stage 2, a large language model reranks the top-$K$ candidates by reading the review and each candidate together. The reviewed paper $q$ is used only to define $\mathcal{C}(q)$ and never as evidence.

\subsection{Candidate Prior}

For each reviewed paper $q$, benchmark construction produces a candidate
set $C(q) \subset U$ of topically plausible submissions. ECHO applies no
additional filtering and assigns all candidates an equal prior, ranking
them solely from textual evidence in the anonymous review, candidate paper,
and rebuttal.

\subsection{Joint Evidence Model}

Let $r$ denote the anonymous review and let $q$ denote the paper being reviewed. 
For each review, Section~\ref{sec:benchmark} constructs a candidate set 
$\mathcal{C}(q)=\{c_1,\ldots,c_n\}$ of plausible submissions. 
Each candidate $c_i=(p_i,b_i)$ consists of a paper $p_i$ and its rebuttal $b_i$. 
Our goal is to rank these candidates according to how likely each submission was authored by the reviewer who wrote $r$.

For candidate $c_i$, let $H_i$ denote the event that $c_i$ is the reviewer's own submission. 
We extract two groups of evidence,
\[
\mathbf{E}_i = E(r,p_i,b_i)
\qquad\text{and}\qquad
\mathbf{W}_i = W(r,b_i),
\]
where $\mathbf{E}_i$ contains content-based signals comparing the review with the candidate paper and rebuttal, while $\mathbf{W}_i$ contains writing-style signals comparing the review with the candidate rebuttal.

The two groups capture complementary information. Content features reflect the reviewer's technical interests and terminology, whereas writing features target more persistent stylistic patterns.

Each linkage signal is standardized relative to the other candidates in the same instance. Let $x_{ik}$ denote candidate $i$'s score under signal $k$, where the signals include semantic similarity, BM25, character $n$-gram
similarity, and authorship-style similarities. We compute
\[
z_{ik}
=
\frac{x_{ik}-\mu_k}
     {\sigma_k},
\]
where $\mu_k$ and $\sigma_k$ are the mean and standard deviation of signal $k$ across all candidates in $\mathcal{C}(q)$. Here, each signal is computed by comparing the anonymous review with the candidate paper or rebuttal using the corresponding semantic, lexical, character-level, or writing-style similarity measure. This normalization measures how strongly a candidate matches the review relative to the other plausible submissions in the same instance.

We combine the standardized signals using a conditional-logit model,
\[
U_i
=
\boldsymbol{\theta}_E^\top \mathbf{z}_{E,i}
+
\boldsymbol{\theta}_W^\top \mathbf{z}_{W,i},
\]
where $\mathbf{z}_{E,i}$ and $\mathbf{z}_{W,i}$ contain the standardized content and writing-style signals for candidate $i$. The conditional-logit likelihood is defined by
\[
P(H_i \mid r,\mathcal{C}(q))
=
\frac{\exp(U_i)}
{\sum_{c_j\in\mathcal{C}(q)}\exp(U_j)}.
\]
The probability is used to fit the fusion coefficients, while candidates are ranked directly by $U_i$. The coefficients $\boldsymbol{\theta}_E$ and $\boldsymbol{\theta}_W$ are fitted jointly with $\ell_2$ regularization. The regularization strength is selected by inner cross-validation, with both inner and outer splits grouped by reviewer.

\subsection{Expertise and Content Evidence}

The first channel captures the reviewer's technical expertise. Although a review necessarily discusses the paper being evaluated, the reviewer also reveals information through the methods they emphasize, terminology they use, assumptions they question, and technical comparisons they introduce.

We compare this information against the content of both artifacts associated with candidate $c_i$. The paper provides the candidate's research topic through its title and abstract, while the rebuttal provides additional technical discussion that is often more detailed and responsive than the abstract alone. Our content channel consists of three families of signals computed between the anonymous review and both $p_i$ and $b_i$, including semantic similarity from pretrained text encoders, lexical similarity from BM25~\citep{robertson2009probabilistic} over unigrams and bigrams, and fine-grained character $n$-gram similarity. These are the three content-signal families used in our model. The semantic family contains multiple scores obtained from different encoders and aggregation strategies. Because reviews and rebuttals are long, semantic similarity is computed over paragraph-level chunks using the best-matching chunk pair, the mean of each review chunk's best candidate match, and the similarity between mean chunk embeddings.

Because candidates already come from the same broad research area, these scores are intended to capture finer-grained expertise than simple field-level similarity. Their within-instance standardization further measures whether a candidate is unusually compatible with the review relative to other plausible submissions in the same candidate set.

\subsection{Writing-Style Evidence}

The rebuttal provides a second source of evidence beyond direct technical-content similarity. Prior work in computational stylometry has shown that authorship can be reflected in persistent patterns such as function-word usage, punctuation, lexical preferences, and syntactic structure \citep{stamatatos2009survey,kestemont2014function}. Reviews and rebuttals share similar forms of scientific argumentation, but directly comparing their raw text can confound writing style with topical content \citep{sundararajan-woodard-2018-style}. 

Figure~\ref{fig:example} illustrates the two types of linkage evidence captured by \textsc{ECHO}, including shared technical terminology and recurring writing-style patterns between a review and its matched rebuttal.

\begin{figure}[t]
  \centering
  \includegraphics[width=0.9\columnwidth]{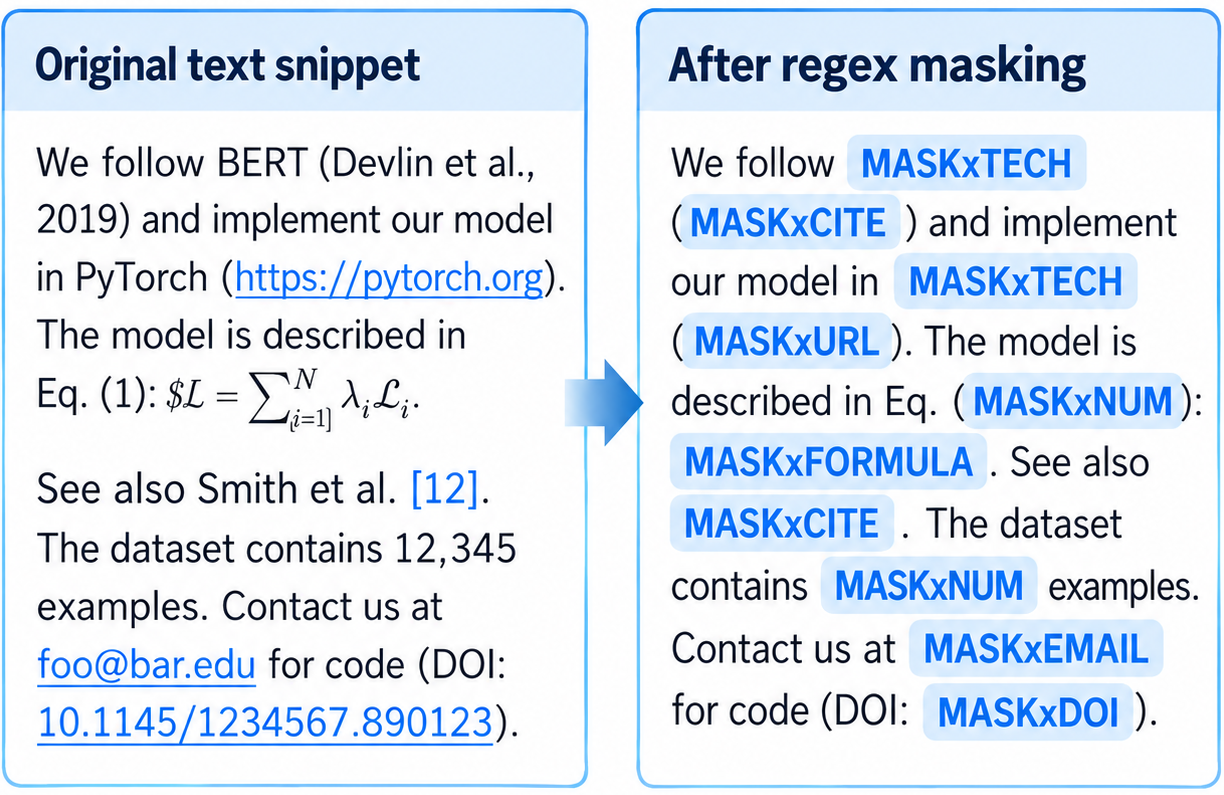}
  \caption{Example of the regex preprocessing stage used before content
  masking. The left panel shows the original text, and the right panel shows
  the text after regex-based masking. Content-bearing items such as citations,
  URLs, formulas, numbers, email addresses, and technical terms are replaced
  with placeholder tokens, while the surrounding structure is preserved.}
  \label{fig:masking_example}
\end{figure}

We therefore construct content-masked versions of the anonymous review and each candidate rebuttal before extracting handcrafted style features. As shown in Figure~\ref{fig:masking_example}, the masking procedure replaces content-bearing tokens with placeholders while preserving function words, pronouns, hedges, discourse markers, punctuation, and sentence structure. We then compute structural and lexical-style features from the masked texts and compare them by cosine similarity after genre-specific standardization.

The writing-style channel additionally includes a pretrained authorship
similarity computed from the raw review and rebuttal texts. Because this representation may retain some topical information, we do not treat the overall channel as strictly content-free. The masked handcrafted signal is designed to reduce topical overlap while
preserving structural and stylistic properties.

Together, the two channels capture complementary aspects of the reviewer.
Content-based evidence reflects \emph{what} the reviewer knows and works on,
using information from both the paper and rebuttal, while the writing-style
channel captures authorship-related linguistic patterns. Their joint
conditional-logit score produces an initial ranking of candidate submissions.

\subsection{LLM Reranking}
Recent work has explored large language models for scientific paper
assessment and automated peer review
\citep{zhou-etal-2024-llm,zhu-etal-2025-deepreview}.
Building on this capability, we use an LLM as a second-stage reranker
to capture finer-grained compatibility between an anonymous review and
a candidate submission. The fusion model first scores each candidate
using fixed similarity measures. We then rerank the top-$K$ candidates
by jointly presenting the review and each candidate to the LLM. For each candidate $c_i$ in this set, the model receives the anonymous review, the candidate's title and abstract, and its rebuttal, and returns a score $s_i\in[0,100]$ indicating how likely candidate $c_i$ was authored by the reviewer, based jointly on technical compatibility between the review and candidate paper and on writing similarities between the review and candidate rebuttal. The prompt asks the model to consider both research expertise and writing habits. No reviewer-identifying information is provided, and the reviewed paper $q$ is not shown.

The LLM score is combined with the fusion score within the top-$K$ set,
\[
U'_i
=
\alpha\, z_K(U_i)
+
(1-\alpha)\, z_K(s_i),
\]
where $z_K(\cdot)$ standardizes a score over the $K$ candidates and $\alpha$ is selected on reviewer-disjoint folds.

The complete prompt for LLM reranking is provided in Appendix~\ref{app:prompt}.
\begin{table*}[t]
\centering
\small
\begin{tabular}{lrrrrrr}
\toprule
\textbf{Method}
& \textbf{Top-1}
& \textbf{Top-5}
& \textbf{Top-10}
& \textbf{Top-20}
& \textbf{Time}
& \textbf{Cost} \\
\midrule

\multicolumn{7}{l}{\emph{Direct full-pool baselines}} \\
Random
& 0.5\%
& 2.6\%
& 5.3\%
& 10.5\%
& --
& -- \\

Open-source LLM (Qwen3-32B)
& 7.0\%
& 14.4\%
& 21.1\%
& 31.5\%
& 14.5 h
& -- \\

Closed-source LLM (GPT-6 Luna)
& 18.5\%
& 35.6\%
& 44.6\%
& 54.7\%
& 4.5 h
& \$40.9 \\

\midrule
\multicolumn{7}{l}{\emph{Primary \textsc{ECHO} configuration}} \\

\textsc{ECHO} (GPT-6 Luna)
& 34.6\%
& 54.0\%
& 63.4\%
& \textbf{67.4\%}
& 2.5 h
& \textbf{\$2.74} \\

\midrule
\multicolumn{7}{l}{\emph{\textsc{ECHO} with stronger Stage-2 rerankers}} \\

\textsc{ECHO} (Claude Opus 5.5)$^\dagger$
& 44.3\%
& 60.4\%
& 64.8\%
& \textbf{67.4\%}
& 3.7 h
& \$81 \\

\textsc{ECHO} (GPT-6 Astra)
& \textbf{56.4\%}
& \textbf{65.4\%}
& \textbf{66.4\%}
& \textbf{67.4\%}
& \textbf{2.4 h}
& \$143 \\

\midrule
\multicolumn{7}{l}{
\emph{Full-pool scoring on a random subset of 20 instances
(9{,}604 candidates)}
} \\

Closed-source LLM (GPT-6 Luna)
& 20.0\%
& 35.0\%
& 35.0\%
& 45.0\%
& 0.3 h
& \$2.8 \\

Closed-source LLM (GPT-6 Astra)
& 70.0\%
& \textbf{95.0\%}
& \textbf{95.0\%}
& \textbf{95.0\%}
& 0.5 h
& \$228 \\

\textsc{ECHO} (GPT-6 Luna)
& 50.0\%
& 55.0\%
& 80.0\%
& 80.0\%
& $<$0.1 h
& \textbf{\$0.18} \\

\textsc{ECHO} (GPT-6 Astra)
& \textbf{75.0\%}
& 80.0\%
& 80.0\%
& 80.0\%
& $<$0.1 h
& \$9.6 \\

\bottomrule
\end{tabular}
\caption{
Main results on R2S-Bench (298 instances, median 414 candidates).
Top-$k$ is the share of instances whose true submission is ranked within
the first $k$. GPT-6 Luna is the primary Stage-2 reranker.
\emph{Time} is wall-clock time and \emph{Cost} is API spend.
The final block reports full-pool scoring on a random subset of 20 instances,
together with the corresponding \textsc{ECHO} results on the same subset.
$^\dagger$Claude Opus 5.5 declined 9.2\% of pairwise judgements due to
its safety guard. Declined judgements are assigned neutral scores.
}
\label{tab:main}

\end{table*}
\section{Evaluations}
\label{sec:experiments}

\subsection{Evaluation Setup}
\label{sec:setup}

\paragraph{Benchmark.}
All experiments use the R2S-Bench described in Section~2, containing 298 instances from 256 unique reviewer--authors. Each instance contains a candidate set with a median of 414 submissions (range 17 to 1{,}181), drawn from a pool of 5{,}061 distinct submissions across all instances. Each candidate is represented by its paper title and abstract together with its cleaned rebuttal.

\paragraph{Metrics.}
We report Top-1, Top-5, Top-10, and Top-20 accuracy, measuring how often the reviewer's own submission appears within the corresponding rank cutoff. We also report the wall-clock time required to score the full benchmark and the corresponding API cost.

\paragraph{Baselines.}
We compare against random ranking and direct LLM-based ranking over the full candidate set. For the latter, we evaluate both a closed-source model and an open-weight model using the same review-to-submission scoring prompt. Retrieval and authorship-matching components used by our method are evaluated separately in the ablation study.

\begin{table}[t]
\centering
\small
\begin{tabular}{lrrr}
\toprule
\textbf{Variant} & \textbf{Top-1} & \textbf{Top-5} & \textbf{Top-10} \\
\midrule

\multicolumn{4}{l}{\emph{Stages}} \\
Fusion only
& 26.5\% & 47.0\% & 58.7\% \\
LLM only on fusion top-20
& 24.8\% & 51.7\% & 62.4\% \\
Fusion + LLM
& \textbf{34.6\%} & \textbf{54.0\%} & \textbf{63.4\%} \\

\midrule
\multicolumn{4}{l}{\emph{Candidate artifacts}} \\
Paper only
& 30.5\% & 49.0\% & 54.7\% \\
Rebuttal only
& 30.9\% & 52.7\% & 59.7\% \\
Paper + rebuttal
& \textbf{34.6\%} & \textbf{54.0\%} & \textbf{63.4\%} \\

\midrule
\multicolumn{4}{l}{\emph{Evidence}} \\
Content only
& 28.9\% & 52.0\% & 59.4\% \\
Writing style only
& 24.8\% & 42.3\% & 48.3\% \\
Content + writing style
& \textbf{34.6\%} & \textbf{54.0\%} & \textbf{63.4\%} \\

\bottomrule
\end{tabular}
\caption{Ablation results on R2S-Bench. Stage ablations compare different
uses of the fusion and LLM scores. Artifact and evidence ablations use
the complete linkage pipeline unless otherwise noted.}
\label{tab:ablation}
\end{table}
\begin{table*}[t]
\centering
\small
\begin{tabular}{lcc|rrr|rr}
\toprule
& \multicolumn{2}{c|}{\textbf{Similarity to original}} &
\multicolumn{3}{c|}{\textbf{Linkage}} &
\multicolumn{2}{c}{\textbf{Bits}} \\
\textbf{Review} & semantic & style &
Top-1 & Top-5 & Top-10 & content & style \\
\midrule
Original      & --    & --   & 34.6\% & 54.0\% & 63.4\% & 1.12 & 0.43 \\
Light rewrite & 0.998 & 0.92 & 22.1\% & 42.3\% & 56.7\% & 1.12 & 0.37 \\
Heavy rewrite & 0.993 & 0.59 & 20.5\% & 40.3\% & 49.7\% & 1.09 & 0.17 \\
\bottomrule
\end{tabular}
\caption{End-to-end linkage performance after AI rewriting. Similarity compares rewritten reviews with the originals using SPECTER2 and masked writing-style similarity. Bits denote information gain over uniform guessing from the content and writing-style channels.}
\label{tab:rewriting}
\end{table*}
\subsection{Main Results}
\label{sec:main_results}

Table~\ref{tab:main} shows that review-to-submission linkage is practically feasible. Even direct LLM scoring exceeds random ranking, while \textsc{ECHO} exploits these signals more effectively. Among a median of 414 plausible candidate submissions, it places the reviewer's own submission in the top 5 for 54.0\% of reviews and in the top 10 for 63.4\%, compared with 2.6\% and 5.3\% by chance. The closed-source LLM reaches only 35.6\% and 44.6\% on the same metrics, despite scoring every candidate directly. Because \textsc{ECHO} applies LLM scoring only to a small shortlist, this gap shows the value of combining content and writing-style evidence before reranking. The staged design is also substantially more efficient, completing the benchmark in 2.5 hours at \$2.74, compared with 4.5 hours and \$40.9 for full-pool closed-source LLM scoring. Stronger rerankers raise the attack accuracy further. With Claude Opus 5.5 or GPT-6 Astra in Stage~2, Top-1 rises from 34.6\% to 44.3\% and 56.4\%, while Top-20 stays at 67.4\% because Stage~1 is unchanged and its recall bounds all three \textsc{ECHO} variants.

Claude Opus 5.5 declined 551 of 5{,}957 pairwise judgements across 42 instances. Declined candidates receive the shortlist mean score, making this row a conservative estimate.

On the 20-instance subset, full-pool GPT-6 Astra reaches 95.0\% Top-10, compared with 80.0\% for \textsc{ECHO} with Astra, but costs \$228 versus \$9.6 for \textsc{ECHO} on the same subset. Extrapolating full-pool scoring to the entire benchmark would cost about \$3{,}400, while \textsc{ECHO} with Astra costs \$143 for the full benchmark.

Overall, these results establish the practical exploitability of review-to-submission linkage. Using only routinely released peer-review artifacts, an attacker can in many cases narrow hundreds of plausible submissions to a small shortlist containing the reviewer's own paper.

\subsection{Ablation Study}
\label{sec:ablation}

\paragraph{Ranking stages.}
The fusion stage already provides a strong shortlist, reaching 47.0\% Top-5 and 58.7\% Top-10. Applying the LLM score alone to the same top-20 shortlist reaches 51.7\% Top-5 and 62.4\% Top-10, showing that the reranker provides additional signal beyond the initial fusion ranking. Combining the fusion and LLM scores performs best, increasing Top-1 from 26.5\% to 34.6\%, Top-5 from 47.0\% to 54.0\%, and Top-10 from 58.7\% to 63.4\%.

\paragraph{Candidate artifacts.}
Both candidate artifacts contribute to linkage performance. Using only the paper yields 54.7\% Top-10 accuracy, while using only the rebuttal reaches 59.7\%. Combining both improves Top-10 to 63.4\% and Top-1 from approximately 31\% with either artifact alone to 34.6\%. The relatively strong rebuttal-only performance suggests that rebuttals carry substantial technical information in addition to writing-style cues.

\paragraph{Evidence channels.}
Content evidence alone reaches 52.0\% Top-5 and 59.4\% Top-10, while the writing-style channel alone reaches 42.3\% and 48.3\%, respectively. Combining the two channels yields the strongest performance across all metrics, increasing Top-1 from 28.9\% with content alone to 34.6\%. This fact shows that the linkage does not depend on a single incidental textual cue. Instead, it is supported by multiple complementary signals that persist across different artifacts.

\subsection{Effect of AI Rewriting}
\label{sec:rewriting}

Rewriting has been studied as a way to suppress authorship and
stylistic signals while preserving the underlying content
\citep{bevendorff-etal-2019-heuristic,
bo-etal-2021-er,
mireshghallah-berg-kirkpatrick-2021-style,
fisher-etal-2024-styleremix}.
We evaluate whether AI rewriting can reduce stylistic cues available to the attacker under two settings. The \emph{light} setting improves grammar and
clarity while largely preserving wording and structure, while the
\emph{heavy} setting paraphrases the review into neutral academic English.
We rerun the full linkage pipeline without adapting the attacker.

Both settings preserve semantic content, with SPECTER2 similarity above 0.99. Content-channel information (information gain over uniform guessing, measured in bits; Appendix~\ref{app:impl}) remains nearly unchanged, decreasing from 1.12 to 1.09 bits under heavy rewriting. In contrast, masked style similarity falls from 0.92 under light rewriting to 0.59 under heavy rewriting, while style-channel information drops from 0.43 to 0.17 bits.

This reduction in stylistic evidence weakens but does not eliminate the
attack. Top-10 decreases from 63.4\% to 56.7\% under light rewriting and
49.7\% under heavy rewriting, still far above the 5.3\% chance level.
These results show that AI rewriting affects stylistic evidence much more
strongly than content-based evidence.

\section{Conclusion}

We introduced \emph{review-to-submission linkage}, a privacy risk in which an anonymous review can be associated with the reviewer's own submission using routinely released peer-review artifacts. We constructed R2S-Bench and developed \textsc{ECHO} to evaluate this threat in practice. Among a median of 414 plausible candidate submissions, our primary configuration places the reviewer's own submission in the top 5 for 54.0\% of reviews and in the top 10 for 63.4\%, while a stronger Stage-2 reranker reaches 56.4\% Top-1 and 66.4\% Top-10. The linkage is supported by complementary content and writing-style signals and remains substantial even after AI rewriting suppresses stylistic cues. These findings show that reviewer anonymity must account not only for individual anonymized documents, but also for links inferred across released peer-review artifacts.
\section*{Limitations}

\paragraph{Benchmark versus conference reviewing.}
Our benchmark is constructed from the F1000 open-research platform because it provides observable reviewer identities for establishing ground truth, but its review process differs from that of large double-blind conferences. F1000 reviewers are invited by editors or authors rather than assigned through conference-style affinity matching, and the platform spans a broad range of disciplines. As a result, our eight research areas provide only a coarse approximation of the topical structure of a single conference. Candidate sets are also smaller than a conference-wide submission pool, with a median of 414 candidates in our benchmark compared with thousands of submissions at venues such as ICLR. Accordingly, absolute linkage rates may not transfer directly across settings; the degree of anonymity-set reduction may provide a more robust measure of risk. In addition, F1000 author responses are signed by individual authors, allowing us to verify that the ground-truth rebuttal was written by the matched reviewer. Conference rebuttals are typically submitted on behalf of the full author team and may therefore provide a weaker writing-style signal.

\paragraph{Applicability to public-review systems.}
Our threat model applies to peer-review systems in which anonymous reviews and candidate submissions are publicly accessible while reviewing is ongoing. Systems that keep either reviews or submissions private until after the review process are outside the scope of our attack. The risk we identify should thus be understood as a limitation of open or partially open review workflows rather than of double-blind reviewing in general.

\paragraph{Shortlist ceiling.}
The LLM reranking stage operates only on the top-$K$ candidates produced by the fusion model. Consequently, its performance is bounded by the recall of the initial shortlist: if the reviewer's submission is absent from the top-$K$, reranking cannot recover it. With $K=20$, the true submission appears in the fusion shortlist for 67.4\% of instances. Improving fusion-stage recall would therefore increase the fraction of instances for which reranking can potentially recover the correct submission.

\section*{Ethical Considerations}

This work studies a potential privacy risk in anonymous peer review. Our goal is to characterize whether routinely released peer-review artifacts can weaken reviewer anonymity and to inform the design of safer review systems, rather than to identify reviewers in ongoing anonymous review processes.

Our benchmark is constructed from the F1000 open-review platform, where reviewer identities and reviews are already publicly available. Reviewer identities are used only to establish ground-truth review-to-submission links and are removed from all linkage inputs. We additionally sanitize reviews to remove explicit identity cues such as names, affiliations, email addresses, ORCID identifiers, profile links, and signatures.

Review-to-submission linkage could nevertheless be misused to narrow the identity of anonymous reviewers. We therefore emphasize the privacy implications of releasing multiple related review artifacts and encourage venues to consider cross-artifact leakage when deciding what information to make public and when to release it. Our results should be interpreted as evidence of a privacy risk in open or partially open review workflows, rather than as a tool for identifying individual reviewers.


\bibliography{custom}

\appendix

\section{Linkage Signals}
\label{app:signals}

\textsc{ECHO} fuses 46 signals, 22 computed between the review and the candidate paper (title and abstract) and 24 between the review and the candidate rebuttal.

\paragraph{Semantic similarity.}
We use five encoder families, with SPECTER2 evaluated under two adapter configurations, yielding six embedding variants in total. These are SPECTER2 (\texttt{allenai/specter2\_base}) with the proximity adapter and, separately, the adhoc-query adapter on the review side, SciNCL, BGE-large-en-v1.5, E5-large-v2, and Qwen3-Embedding-8B. Reviews and rebuttals are split into paragraph-level chunks of 120--200 words. Each embedding variant produces three aggregation scores, using the best review--candidate chunk pair, the mean over review chunks of each chunk's best candidate match, and the cosine similarity between mean chunk embeddings. This gives 18 chunk-level signals per artifact. Qwen3-Embedding-8B additionally encodes the full text in one pass, yielding 19 semantic signals per artifact and 38 in total. Qwen3-Embedding-8B is the strongest single encoder, reaching 43.6\% Top-10 against the rebuttal, compared with 36.6\% for SPECTER2.

\paragraph{Lexical similarity.}
We use BM25 over content words, once with unigrams and once with unigrams plus bigrams, and cosine similarity of character 2--4-gram TF-IDF vectors (sublinear term frequency, minimum document frequency 3, at most 200,000 features). Each is computed against the paper and the rebuttal, giving 6 signals.
\paragraph{Writing style.}
We use two rebuttal-side signals: the handcrafted style similarity of Section 3.4, computed on content-masked text, and the STAR authorship encoder (AIDA-UPM/star) on the raw rebuttal. Abstracts are short, formulaic and multi-authored, so no style signal is computed against the paper. We also tested LUAR, LUAR-CRUD and StyleDistance. None improved the fusion once STAR was included, and all three lose most of their accuracy on masked text, which indicates that their signal is largely topical. The handcrafted features are the strongest content-free signal. On masked text they reach a Top-10 of 19.5\%, against 5.3\% by chance, whereas SPECTER2 on the same masked text falls to 7.7\%.

\section{LLM Reranking and Candidate Shortlist}
\label{app:rerank}

\paragraph{Shortlist size.}
The reranker only reorders the fusion top-$K$, so its reachable accuracy is the fusion recall at $K$: 58.7\% at $K=10$, 67.4\% at 20, 70.1\% at 30, 75.8\% at 50 and 81.5\% at 100. The number of LLM judgements, and therefore the cost, grows linearly with $K$ (\$2.7 for GPT-6 Luna at $K=20$ and \$6.9 at $K=50$). We use $K=20$ because recall grows slowly beyond it.

\paragraph{Reranker design.}
Each (review, candidate) pair is scored independently with an integer from 0 to 100 (Appendix~\ref{app:prompt}). Two alternatives performed worse. A Yes/No formulation that reads the probability of \emph{Yes} from token log-probabilities collapses to nearly constant answers once extended reasoning is enabled. For these development comparisons, we additionally report mean reciprocal rank (MRR). A listwise formulation that shows all 20 candidates in one prompt and asks for a full ranking is cheaper but no better than the fusion alone (MRR 0.370 versus 0.364 with DeepSeek-V4-Pro). Reasoning matters: DeepSeek-V4-Pro without reasoning adds nothing as a pointwise reranker (MRR 0.364), whereas GPT-6 Luna with a 2{,}000-token reasoning budget raises MRR to 0.445.

\paragraph{Combination with the fusion score.}
The final score is $\alpha\,z_K(U_i) + (1-\alpha)\,z_K(s_i)$ with $\alpha$ chosen from $\{0, 0.1, \ldots, 1\}$ on reviewer-disjoint folds. The selected value is $0.5$ for GPT-6 Luna and Claude Opus 5.5, and $0$ to $0.1$ for GPT-6 Astra, whose scores alone dominate. Using the LLM score without the fusion score gives a Top-1 of 24.8\% for Luna, 36.9\% for Opus 5.5 and 57.0\% for Astra, so the fusion score still adds information for the two weaker rerankers.
\section{Reviewer--Author Matching and Benchmark Construction}
\label{app:identity}

\paragraph{Identity matching.}
Reviewers are matched to article authors in three tiers, from strongest to weakest, and each reviewer--author pair keeps its strongest tier.
\begin{enumerate}
\item \emph{ORCID}: the review contributor's ORCID equals an article author's ORCID (1{,}539 reviewer--authors).
\item \emph{ORCID works}: the reviewer's public ORCID works list contains the article, and the article has an author with the same surname and first initial (59).
\item \emph{Name and affiliation}: identical full given name and surname, the name is not carried by more than one ORCID in the corpus, there is no conflicting ORCID on either side, and the reviewer and author share at least two affiliation tokens, at least one of which is distinctive, meaning it occurs in fewer than 1\% of all affiliation strings (1{,}129).
\end{enumerate}
A fourth tier, identical full name without affiliation evidence, matches 454 further people but is never used, because common names make it unreliable. The three usable tiers give the 2{,}727 reviewer--authors reported in Section~2.1, out of 26{,}144 distinct reviewers (14{,}252 with an ORCID) and 53{,}491 distinct authors. Every matched person receives a pseudonymous identifier, and all ORCIDs, names and affiliations are discarded before benchmark construction.

\paragraph{Candidate sets.}
For a review of paper $q$ written on date $t$, the candidate set contains every article that belongs to the same research area as $q$, was publicly available by $t$ plus 182 days, and has an accompanying rebuttal. After selecting the ground-truth submission, we remove $q$ itself, every other article authored by the matched reviewer under any of the three tiers, and every article with an author whose name could plausibly be the reviewer, including the same name key or the same surname and first initial with an unresolved ORCID. The 182-day allowance reflects that, at a conference, submissions and rebuttals in the same cycle become visible together.

\paragraph{Ground truth.}
The ground-truth submission is one article authored by the reviewer within the same research area and time window, with a rebuttal written by the matched reviewer, in English, of at least 50 words. Rebuttals are cleaned by removing quoted review text, signatures and explicit identity information. When a reviewer has several eligible articles, one is chosen at random, and at most two reviews per reviewer enter the benchmark. These constraints leave 298 instances from 256 reviewer--authors. By matching tier, the ground-truth link is established through ORCID for 233 instances, through ORCID works for 12, and through name and affiliation for 53 (201, 9 and 46 of the 256 reviewer--authors, respectively).

\section{Review Sanitization Details}
\label{app:sanitize}

F1000 reviews are signed and embedded in a platform template, so each review is cleaned before it is used as a query. The \emph{main} version used in all experiments applies the following steps.

\begin{enumerate}
\item Remove template checklist questions, identified as question blocks that recur in at least 20 reviews, together with their one-word answers (\emph{yes}, \emph{no}, \emph{partly}, \emph{not applicable}).
\item Remove the platform confirmation statement (``I confirm that I have read this submission \ldots'') and the ``Reviewer Expertise'' field.
\item Remove signature blocks: trailing short paragraphs containing a sign-off or a contributor name.
\item Replace reviewer and co-reviewer names with \texttt{[NAME]}, e-mail addresses with \texttt{[EMAIL]}, ORCID identifiers with \texttt{[ORCID]}, profile links (ORCID, ResearchGate, LinkedIn, Google Scholar, social media) with \texttt{[PROFILE-URL]}, and affiliation strings with \texttt{[AFFILIATION]}.
\end{enumerate}

A \emph{strict} version additionally masks all URLs, DOIs and PubMed identifiers, citations of the reviewer's own name, titles of the reviewer's own papers, and sentences that refer to the reviewer's own work. Manual inspection of 500 sanitized reviews found no remaining names, e-mail addresses, ORCID identifiers, profile links or expertise statements. Rebuttals are cleaned in the same way, with quoted review text additionally removed.

\section{LLM Reranking Prompt}
\label{app:prompt}

Every judgement uses the system prompt below and a user message containing the review, the candidate's title and abstract, and its rebuttal, truncated to 1{,}500, 300 and 1{,}200 words respectively. No identity information and no text of the reviewed paper are included. The full-pool LLM baselines use the same prompt for every candidate.

\begin{quote}\small\ttfamily\raggedright
\textbf{System.} You are an expert in authorship analysis of scientific writing. You compare an anonymous peer review with a candidate submission (its abstract and the author response its authors posted during review) and judge whether the review and the author response were written by the same researcher. Consider research expertise and interests revealed by the texts, and writing habits: sentence structure, punctuation, hedging, discourse markers, politeness formulas, spelling conventions, how points are organised. Reply with a single integer from 0 to 100 giving the likelihood that they were written by the same person.

\medskip
\textbf{User.}\\
=== ANONYMOUS PEER REVIEW (written by reviewer R about some other paper) ===\\
\{review\}\\[2pt]
=== CANDIDATE SUBMISSION ===\\
Title: \{title\}\\
Abstract: \{abstract\}\\[2pt]
=== AUTHOR RESPONSE posted by the candidate submission's authors ===\\
\{rebuttal\}\\[2pt]
Question: How likely is it that the AUTHOR RESPONSE was written by the same researcher who wrote the ANONYMOUS PEER REVIEW? Reply with a single integer from 0 (certainly a different person) to 100 (certainly the same person), and nothing else.
\end{quote}

\section{Implementation Details}
\label{app:impl}

\paragraph{Fusion and evaluation.}
Signals are standardized within each candidate set and combined by a conditional logit with $\ell_2$ regularization. Evaluation is 5-fold cross-validation grouped by reviewer. The regularization strength is chosen from $\{0.01, 0.1, 1, 10\}$ by an inner reviewer-grouped cross-validation inside each training fold. All reported numbers are out-of-fold.

\paragraph{Content masking.}
Masking uses spaCy (\texttt{en\_core\_web\_sm}) part-of-speech and entity tags after the regex pre-processing of Figure~3, with word frequencies from \texttt{wordfreq}. About half of all tokens are masked, including every noun, proper noun, adjective and number, and verbs and adverbs outside the 5{,}000 most frequent English words unless they are hedges or discourse markers. Every mask tag is rendered as the same placeholder before feature extraction, so the tag type carries no information. The handcrafted representation has 122 features: sentence- and word-length statistics, type--token ratio, rates of 17 punctuation marks, frequencies of 85 function words, hedge and discourse-marker rates, first-person pronoun rates, question rate, contraction rate, British versus American spelling rates, and upper-case and digit character fractions. Features are z-scored within genre (reviews against reviews, rebuttals against rebuttals) and compared by cosine similarity.

\paragraph{LLM access and cost.}
GPT-6 Luna in ECHO was queried with a 2,000-token reasoning budget and
a 2,400-token output limit. Claude Opus 5.5, GPT-6 Astra, and GPT-6 Luna
in the full-pool baseline were queried with low reasoning effort and the
same output limit. Costs are computed from measured prompt lengths and
reported output tokens using the corresponding model list prices.
The open-weight baseline runs Qwen3-32B with vLLM on a single 96\,GB GPU,
with thinking disabled and a 16-token output limit.

\paragraph{Declined judgements.}
Claude Opus 5.5 returned an empty answer with stop reason \texttt{content\_filtered} for 551 of the 5{,}957 judgements, affecting 42 instances, 31 of them entirely. Repeated requests recovered only a small fraction, so these are treated as refusals by the model's safeguards. A declined candidate receives the mean LLM score of its shortlist, which leaves the fusion order unchanged for instances declined entirely. The OpenAI models declined no judgements.

\paragraph{Subset for full-pool scoring.}
The 20 instances in the third block of Table~2 are the first 20 of a uniformly random permutation of all 298 instances. They come from 20 distinct reviewers and five research areas, with candidate sets ranging from 33 to 1{,}050 submissions (9{,}604 in total). \textsc{ECHO} results in that block are the full-benchmark runs restricted to these instances.

\paragraph{AI rewriting.}
Reviews are rewritten by GPT-6 Luna. The \emph{light} prompt asks the model to correct grammar, spelling and punctuation and improve clarity while keeping the wording, sentence order, paragraph structure and every point unchanged. The \emph{heavy} prompt asks it to rewrite the review in its own words in neutral, formal academic English, preserving every point, claim, question and request in the same order but changing phrasing, sentence structure and word choice throughout. The attacker is not adapted to rewritten text: fusion weights and $\alpha$ are those fitted on original reviews.

\paragraph{Channel information.}
We quantify the information carried by each evidence channel using out-of-fold conditional-logit probabilities. For each instance, channel scores are standardized within the candidate set and converted to a probability of the true candidate using a conditional logit model whose coefficient is fitted with reviewer-disjoint 5-fold cross-validation. We then compute
\[
I = \log_2 N + \log_2 P(c^\star),
\]
where \(N\) is the candidate-set size and \(P(c^\star)\) is the predicted probability of the true candidate. Uniform guessing gives \(I=0\) bits; larger values indicate more identifying information.

\end{document}